\RequirePackage{fix-cm}
\pdfoutput=1
\documentclass[a4paper,DIV15,10pt,ngerman,twocolumn]{scrartcl}
\usepackage{graphicx}
\usepackage{amsmath}
\usepackage{amssymb}
\usepackage{subfig}
\usepackage{authblk}

\usepackage{color}

\begin{document}

\title{Rheology of weakly wetted granular materials 
- a comparison of experimental and numerical data}

\author[*]{R\"udiger Schwarze}
\author[*]{Anton Gladkyy}
\author[*]{Fabian Uhlig}
\author[**]{Stefan Luding}

\affil[*] {
  Institute of Mechanics and Fluid Dynamics, TU Bergakademie Freiberg.
  Lampadiusstr. 4, 09596 Freiberg, Germany
}

\affil[**] { 
  Multi Scale Mechanics (MSM), Engineering Technology
  (CTW) and MESA+, University of Twente.
  P.O.Box 217, 7500 AE Enschede, The Netherlands
}


\maketitle

\abstract{
Shear cell simulations and experiments of weakly wetted particles 
(a few volume percent liquid binders) are compared, 
with the goal to understand their flow rheology.
Application examples are cores for metal casting by core shooting
made of sand and liquid binding materials.

The experiments are carried out with a Couette-like rotating viscometer. The weakly 
wetted granular materials are made of quartz sand and small amounts of Newtonian liquids. 
For comparison, experiments on dry sand are also performed with a modified configuration 
of the viscometer. The numerical model involves spherical, monodisperse particles with 
contact forces and a simple liquid bridge model for individual capillary bridges between two 
particles. Different liquid content and properties lead to different 
flow rheology when measuring the shear stress-strain relations.
In the experiments of the weakly wetted granular material,
the apparent shear viscosity $\eta_g$ scales inversely proportional to 
the inertial number $I$, for all shear rates. 
On the contrary, in the dry case, an intermediate scaling regime inversely 
quadratic in $I$ is observed for moderate shear rates. 
In the simulations, both scaling regimes are found for dry and wet granular material as well.

}

\section{Introduction}

Dry granular matter and its flow rheology have been the subject 
of detailed studies during the last years and slowly their interesting
behavior becomes more and more clear \cite{GdR04,nakagawa09}. 
The other extreme case are particles suspended in fluids -- a field of 
wide relevance in industrial processes -- which nowadays 
are mostly understood and can be modeled reasonably 
well \cite{Zhu07,hoef08,kafui11,Voi11,Moo11}.

However, weakly wetted granular materials have recently attracted new attention, 
see e.g.\ Ref.\ \cite{Man12,Har2013,Liu2013,Hsiau2013,Zak13},
even though they were studied earlier 
\cite{Lia93,Wei99,Wil00,Her05}. Wet as well as dry granular rheology 
\cite{Jop06,Godd06} plays an important role in geotechnical and geophysical
context \cite{Gab12,Hay11}, as well as in several technical processes
e.g.\ in growth agglomeration \cite{pietsch}, for die-filling \cite{guo11,guo11b},
or in the production of sand cores for casting by core shooting \cite{Bee01}. 
In the latter example, weakly wetted granular materials are mixtures of a 
granular matter and few volume-percent of liquid binder. Like dry
granular materials, they exhibit non-Newtonian flow behavior, where
the relations between shear stresses and shear rates, for example,
can be expressed by nonlinear functions. 
Since the presence of small amounts of liquid change
the rheological behavior of the granular material markedly \cite{Sch08,Lia10},
detailed knowledge of the constitutive equations of these materials
is of fundamental importance for the control of the corresponding
processes. As an example, the filling flow -- even in complex core boxes --
can be analyzed by CFD simulations \cite{Rud09} when the rheological
model of the material is known.

\begin{center}
\begin{figure}[h]
\begin{centering}
\includegraphics[width=0.48\textwidth]{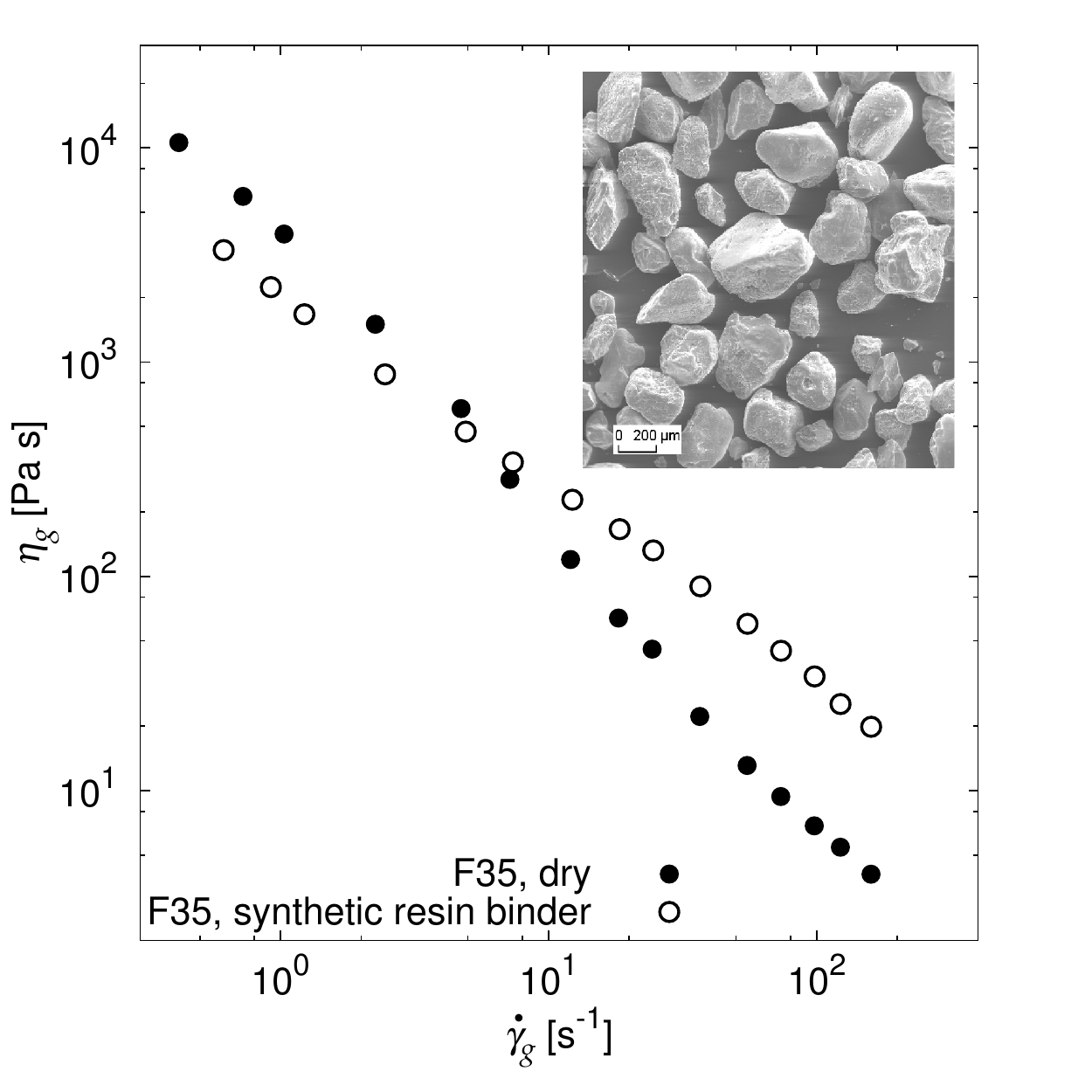}
\end{centering}

\caption{
Characteristics of core shooting material with quartz sand F35 \cite{F35}: 
Rheological data from measurements with dry F35 and with
core shooting material with a mass ratio $m_{SR}/m_{F} =0.02$ between 
the mass $m_{SR}$ of the synthetic resin binder and the 
mass $m_F$ of F35, see section \ref{sec:2} and Ref.\ \cite{Sch08} for 
more details. Inset: SEM picture of a sample of F35.
\label{fig:01}}
\end{figure}

\par\end{center}

As an example, Fig.\ \ref{fig:01} shows some rheometer experiment \cite{Sch08}
flow curves of a typical core shooting material made of quartz sand F35 
(with mean particle diameter $d_{P}=0.18$ mm) \cite{F35} and synthetic
resin binder and the pure, dry F35.
The scaling $\eta_g\left(\dot{\gamma}_g\right)$ of the apparent viscosity
$\eta_g$ with the shear rates $\dot{\gamma}_g$ differs markedly between the
core shooting material and the dry granular F35.

Heuristically, this difference can be explained
by the capillary bridges of the liquid binder between individual sand
particles. A more detailed review of the rheological measurements is 
given in the next section.

Unfortunately, the realization of rheometer experiments involving
weakly wetted granular materials is complicated and time consuming. 
Therefore, alternative and more efficient methods for rheological 
investigations are highly desirable. 
In this paper, we use the split-bottom ring shear setup of a 
rheometer \cite{Fen04} for discrete-element method (DEM) simulations of 
these partly wet granular materials. 
Similar DEM simulations of wet granular
materials have been performed in order to study the micro-mechanics
in cohesive mixing processes \cite{McC03}, the discharge from hoppers
\cite{Ana10} or the mixing in a blade mixer \cite{Rad10}. In these
papers, explicit capillary forces are added to the contact forces
in order to properly describe the interactions between two particles
in the wet granular material. As an alternative to the most simple
approach pursued below, Grima and Wypych \cite{Gri11} employ an implicit
model of the capillary force in their simulations, while 
Mani et al.\ \cite{Man12} explicitly allow for liquid migration between 
the particles and across the bridges.

Starting from the DEM results, we apply a local micro-macro transition 
\cite{Lud08,Lud08Pa,Lud11}
in order to obtain rheological flow rules for weakly wetted granular materials.
First results indicate, that the numerically determined flow rules
exhibit similar differences between dry and wet granular materials
as in the experiments. The device used to measure the 
stress-strain relations is the split-bottom ring-shear cell as invented 
by Fenistein et al.\ \cite{Fen04} and used by others for dry, frictional and
(van der Waals) elasto-plastic adhesive particle systems 
\cite{Lud08,Lud08Pa,Lud11,Sadr08,Bor11,Dij11,Bor12,Wang12,Slot12}.

\section{Previous rheological measurements\label{sec:2}}

Rheological data of weakly wetted granular materials
(core shooting materials) have been recently measured in a Searl-type
rotating viscometer \cite{Sch08} with a fixed bottom
and outer cylinder wall and a rotating inner cylinder wall. Sand particles
are glued to the fixed and rotating cylinder walls in order to define
proper wall shear stress conditions and reduce wall-slip. 
In these shear cell experiments, a shear band width $w_{sb}$ of about 
10 - 20 particle diameters $d_P$ was expected.

For measurements on dry sand,
the gap width $b_{dry}=2$ mm $\simeq 10\,d_P \simeq w_{sb}$ 
between the inner and outer cylinder
was fitted to the assumed width of the shear zone in the granular
material. With this setup, all the material in the gap should be sheared.
Here, the upper annulus of the viscometer gap was open. 

For measurements on core shooting materials, the upper annulus 
was closed by a movable circular ring
in order to superimpose an external pressure $p_{e}$ to the weakly wetted granular
material. Without $p_{e}$, fissures arose in the shear zone, which
induced the disruption of the measurements. In all measurements, $p_{e}$ exceeded the pure
hydrostatic pressure level $p_{h}=\rho\, g\, h\simeq1$\,kPa of the
core shooting material, with bulk density $\rho\simeq1400$ kg/m$^{3}$
and fill level $h\simeq70$ mm in the viscometer gap. 
The gap width $b_{wet}=10$ mm $\simeq 50\,d_P \simeq 5\,w_{sb}$
between the inner and outer cylinder was increased for these experiments. 
We assume, that parts of the material in the gap were not 
sheared in this configuration. Therefore, the exact width of the shear zone remains
unknown, which induces an uncertainty in the absolute values of the shear 
rate. However, the scaling $\eta_g\left(\dot{\gamma}_g\right)$ should have been
correctly resolved in these measurements, too, 
only amplitudes of $\eta_g$ are maybe somewhat shifted.

\begin{figure}
\begin{centering}
\includegraphics[width=0.465\textwidth]{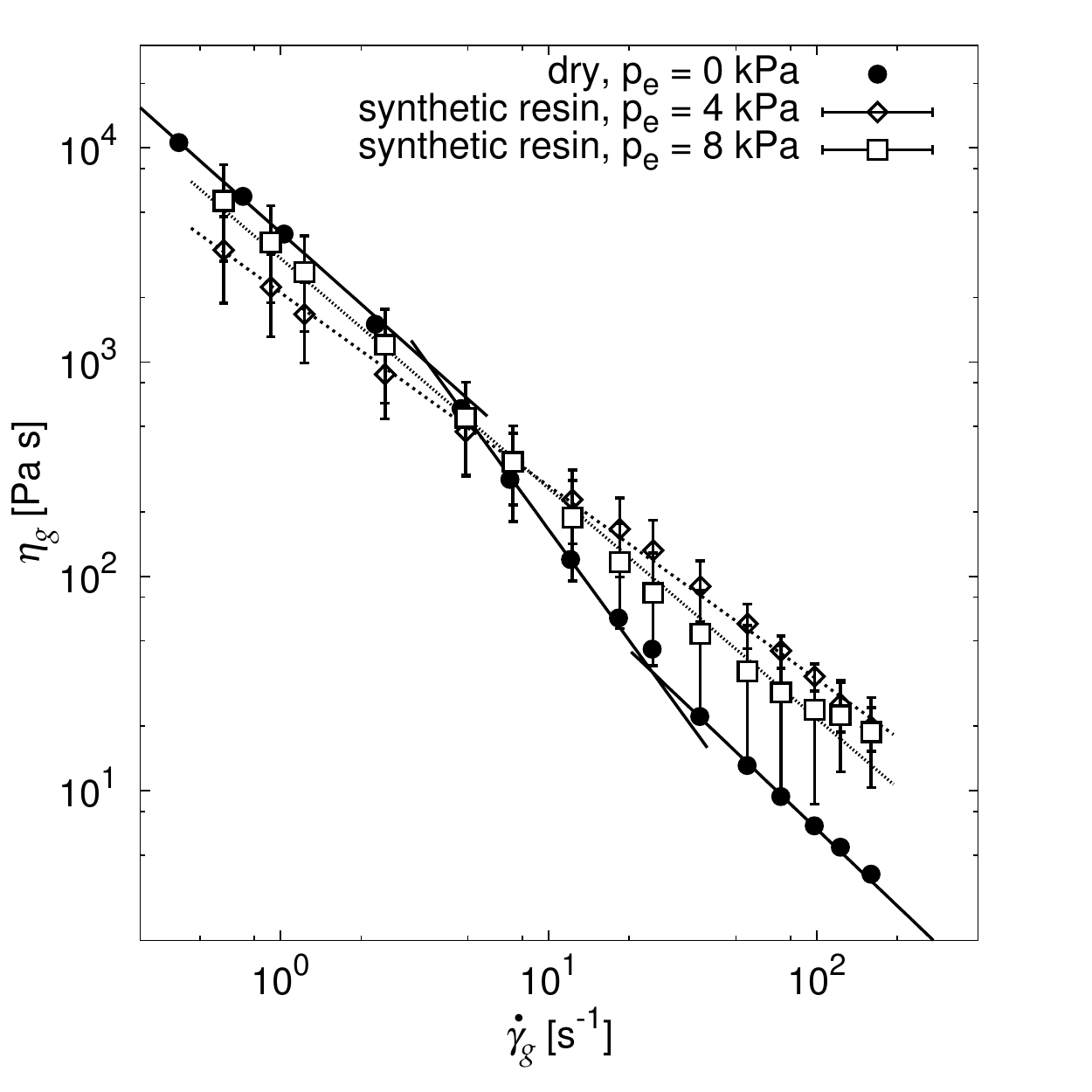}
\par\end{centering}

\begin{centering}
\includegraphics[width=0.465\textwidth]{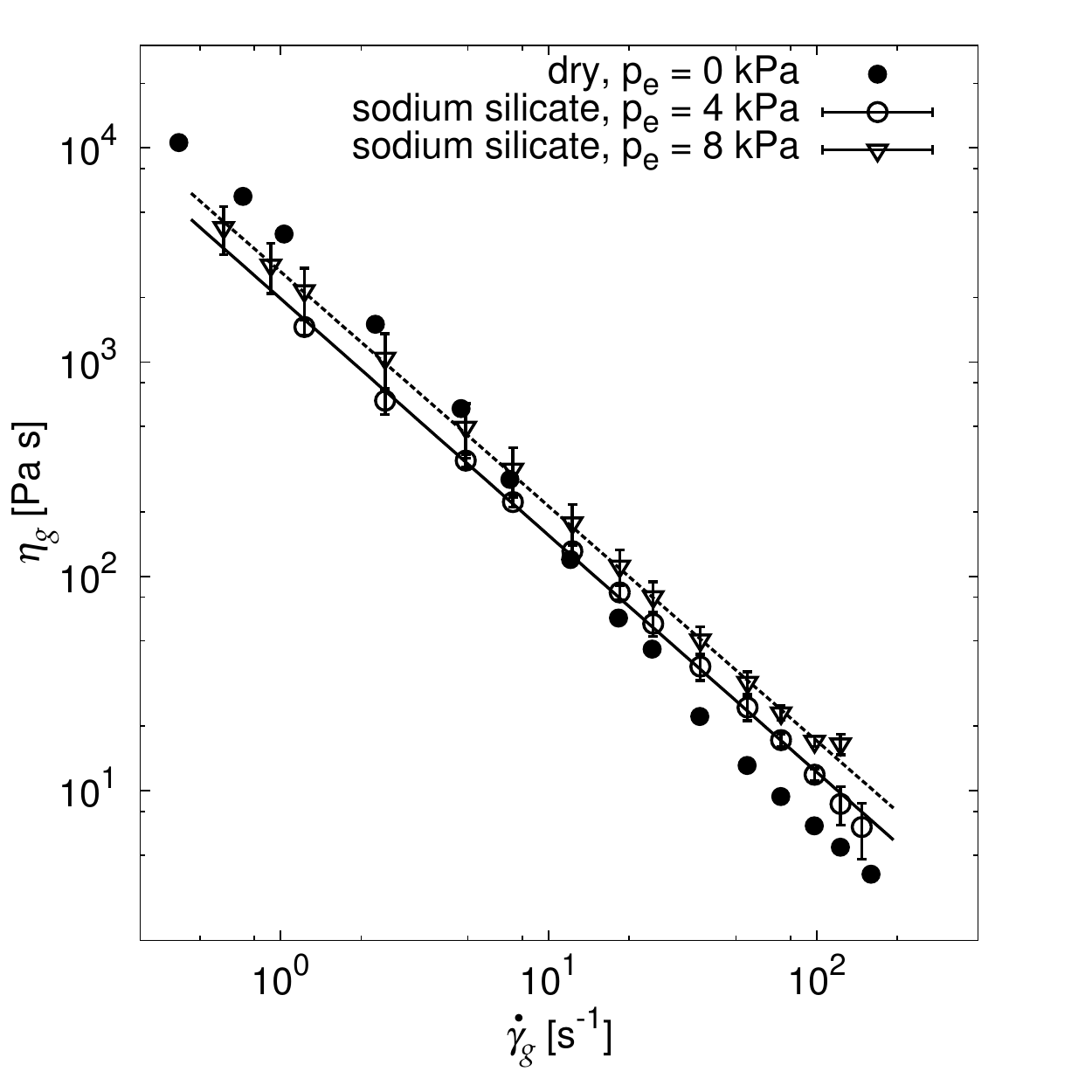}
\par\end{centering}

\caption{Apparent shear viscosity $\eta_g\left(\dot{\gamma}_g\right)$ for core 
shooting material made of F35 and synthetic resin binder with 
$m_{SR}/m_{F} =0.01$ (top), and made of F35 and sodium silicate binder with
$m_{SS}/m_{F} =0.02$ (bottom);
error bars indicate the standard deviation of the experimental data.
Results of dry F35 are also given for comparison, performed with smaller gap $b$. 
Measurements are made with different external pressure levels $p_{e}$. 
Continuous and dashed lines indicate the scaling between $\eta_g$ and $\dot{\gamma}_g$,
obtained from least square fits of Eq.\ (\ref{eq:02}) to the wet data, 
for $0.4$\,s$^{-1} < \dot\gamma_g < 200$\,s$^{-1}$.
}
\label{fig:02}

\end{figure}

In Fig.\ \ref{fig:02}, some results of the rheometer measurements
on two core shooting materials with F35 as basic sand but different liquid binders
are displayed. For both materials, the apparent shear viscosity 
\begin{equation}
\eta_g = \frac{\left|\tau_g\right|}{\dot{\gamma}_g}
\label{eq:01}
\end{equation}
of the core shooting material exceeds
the values of the dry granular F35 markedly for shear rates $\dot{\gamma}_g>10$\,s$^{-1}$.
Here, the quantities $\tau_g$ and $\dot{\gamma}_g$
are \emph{globally} defined, i.e.\ they describe the dynamics of the 
bulk granular material and set-up: 
The mean shear stress $\tau_g=F_{t,o}/A_{o}$ is the quotient of the tangential force 
$F_{t,o}$ at the outer cylinder wall, which is measured by a force transducer, and 
the outer cylinder wall area $A_{o}$. The mean shear rate $\dot{\gamma}_g= U_{i}/b$ is 
approximated by the quotient of the velocity $U_{i}$ of the rotating inner cylinder wall 
and the gap widths $b=b_{dry}$ or $b=b_{wet}$.

Problems in the realization of the measurements are indicated
by the large uncertainty in the results for the core shooting material
with synthetic resin binder and $p_{e}=8$\,kPa.

\begin{table}
\begin{centering}
\begin{tabular}{c|c|c|c|c|}
\cline{2-5} 
 & \multicolumn{2}{c|}{F35, synthetic resin} & \multicolumn{2}{c|}{F35, sodium silicate}\tabularnewline
 & \multicolumn{2}{c|}{$m_{SR}/m_{F} =0.01$ } & \multicolumn{2}{c|}{$m_{SS}/m_{F} =0.02$ }\tabularnewline
\hline 
\multicolumn{1}{|c|}{$p_e$\,[kPa]} & $4$ kPa & $8$ kPa & $4$ kPa & $8$ kPa\tabularnewline
\hline 
\multicolumn{1}{|c|}{$K$} & 2084 & 3291 & 1807 & 2603\tabularnewline
\hline 
\multicolumn{1}{|c|}{$\alpha$} & 0.90 & 1.07 & 1.11 & 1.10\tabularnewline
\hline
\end{tabular}
\par\end{centering}

\caption{Parameters of the flow rules for core shooting materials of 
Fig.\ \ref{fig:02} in Eq.\ (\ref{eq:02}). Values of $K$ and $\alpha$ 
are obtained from a least square fit of Eq.\ (\ref{eq:02}) to the wet data, 
for $0.4$\,s$^{-1} < \dot\gamma_g < 200$\,s$^{-1}$. Here, values of $K$ are mere
fit-parameters since their units are non-linearly dependent 
on the values of $\alpha$ and are thus not given.
}
\label{tab:1}
\end{table}

The flow rules of the core shooting materials in the measured shear 
rate intervals can be described as a first approximation by a power law 
\begin{align}
\eta_g\left(\dot{\gamma}_g\right) & =  K\,\left|\dot{\gamma}_g\right|^{-\alpha}
\label{eq:02}
\end{align}
with the consistency factor $K$ and a power law index $\alpha$. 
(Note that the nonlinear power-law spoils the units and has to
be phrased better using dimensionless quantities as discussed below.)
Table \ref{tab:1} summarizes the values of $K$ and $\alpha$ for the different 
core shooting materials for different levels of $p_{e}$. 

All measurements indicate shear-thinning behavior of the core-shooting
materials with power laws close to $\left(\dot\gamma_g\right)^{-1}$. 
However, some interesting differences are also found: 
At the same levels of $p_{e}$, the parameters $K$ and $\alpha$ differ markedly
between the two core shooting materials with the same basic sand but 
different liquid binders. Obviously, the presence of small amounts of 
different liquids in the same basic granular material affects the flow 
behavior significantly. 
For increasing $p_{e}$, the power law index $\alpha$ 
in case of core shooting material made of F35 with synthetic resin binder decreases, 
whereas $\alpha$ rises for core shooting material made of F35 with sodium silicate binder.

The proposed scaling $\eta_g\propto p_{e}/\dot{\gamma}_g$
of Jop et al. \cite{Jop06} is only found in the core
shooting material with sodium silicate binder, where the continuous
and dashed curves are nearly parallel. In case of the core shooting
material with synthetic resin binder, where the continuous and dashed 
curves intersect, the scaling seems to be more complex.

The rheological behavior of dry sand in the same rheometer 
(but with smaller gap $b$) seems to be 
more complex; we fail to fit the experimental data by a simple power 
law ansatz, Eq.\ (\ref{eq:02}), with constant $K$ and $\alpha$. Instead,
we found three distinct regimes for low, medium and high shear rates. 
The scaling is $\eta_g \propto \dot{\gamma}_g^{-1.1}$ for low, 
$\eta_g \propto \dot{\gamma}_g^{-1.7}$ for medium, and 
$\eta_g \propto \dot{\gamma}_g^{-1.2}$ for high shear rates, 
see Fig.\ \ref{fig:02}(top). 
These different regimes are not found for the weakly wetted 
granular materials.

On the way to a better theoretical description of this 
complex rheology, it is advisable/necessary
to use a dimensionless shear rate:
\begin{align}
I_g & =\frac{\dot{\gamma}_g\, d_{P}}{\sqrt{{p_g}/{\rho}}}
\label{eq:03}
\end{align}
as defined in \cite{GdR04}. 
This number describes the ratio of inertial and confining stress, 
i.e.\ pressure, of the sheared material. For the evaluation of the different
measurements, the global pressure level $p_g$ is estimated by $p_g\simeq p_{e}$,
because $p_{e}\gg p_{h}$ as explained above. The quantities
$\dot{\gamma}_g$, $p_g$, and $I_g$ are \emph{global} parameters,
i.e.\ are obtained from an external measurement that implies
an average over the whole shear cell.
In case of the dry sand, where no external pressure $p_e$ was applied,
$I_g$ cannot be properly evaluated for these measurements.
\begin{figure}
\begin{centering}
\includegraphics[width=0.465\textwidth]{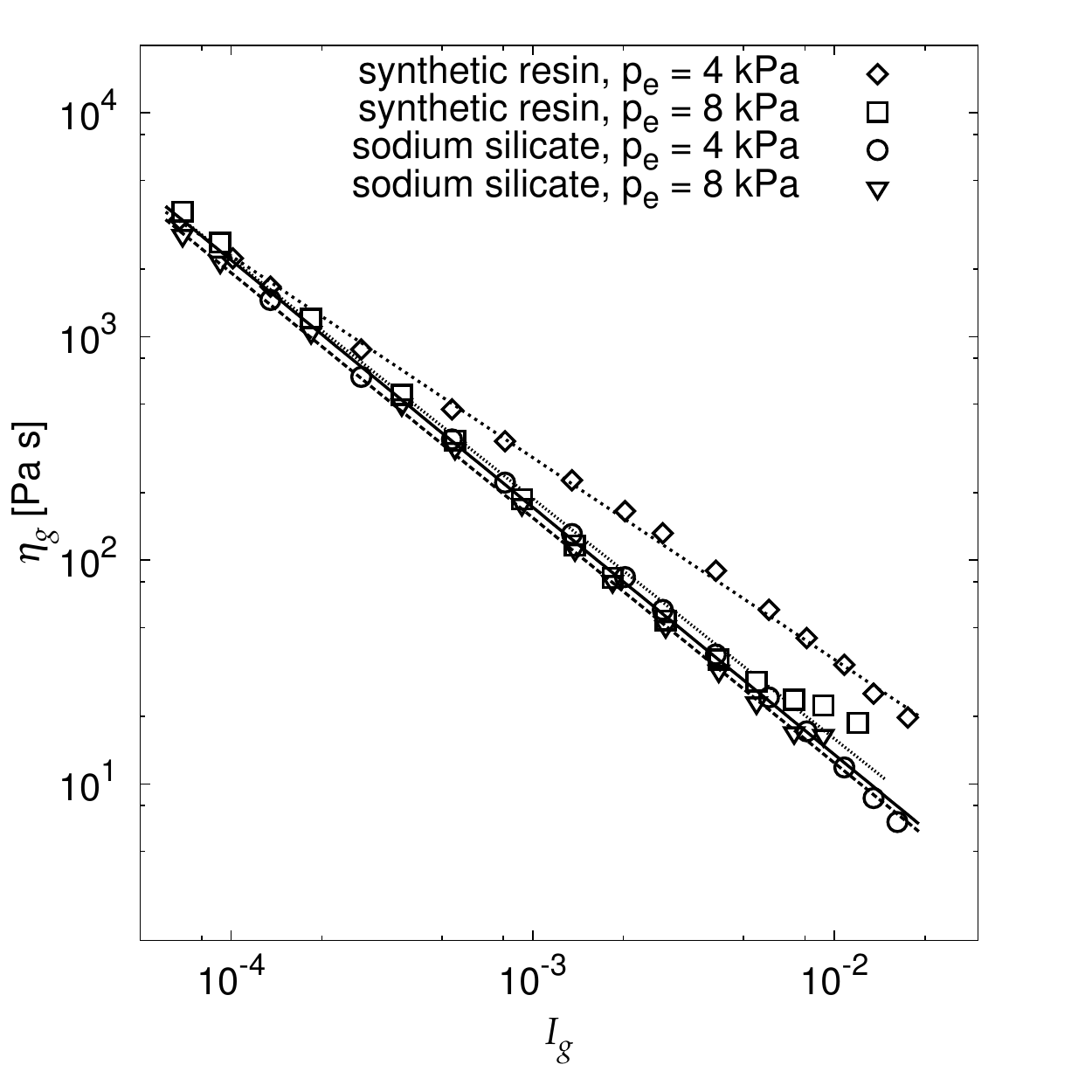}
\par\end{centering}

\caption{Apparent shear viscosity as a function of dimensionless shear 
rate for the two core shooting materials and the two external pressure
levels $p_{e}$ of Fig.\ \ref{fig:02}. 
Continuous and dashed lines indicate the scaling $\eta_g \sim I_g^{-\alpha}$, 
with $\alpha$ from table \ref{tab:1}.
\label{fig:03}}
\end{figure}

Fig.\ \ref{fig:03} summarizes the observed apparent shear viscosities
$\eta_g\left(I_g\right)$ for the different core shooting materials and
the different levels of $p_{e}$. In this semi-dimensionless representation,
all measurements follow roughly
the same shear-thinning behavior
of the core-shooting materials with scaling as
$\eta_g \propto I_g^ {-\alpha}$ with $\alpha \simeq 1$,
which are indicated by continuous and dashed lines as in Fig.\ \ref{fig:02}.

Interestingly, the synthetic binder is sensitive to the confining stress,
whereas the sodium silicate binder is not. The former displays considerably
higher apparent viscosity for smaller $p_e$. The more systematic investigation of 
these subtle differences are ongoing and will be published elsewhere.

\section{Model fundamentals}

The flow of a sheared, weakly wetted granular material is now 
investigated by means of DEM simulations. 

\subsection{Equations of motion}

The DEM model is based on Newton's equations of motion for the translational
and rotational degrees of freedom of a spherical particle
\begin{align}
m_{i}\dfrac{d^{2}\underline{r}_{i}}{dt^{2}} 
& =\sum\limits _{\left\{ i,j\right\} =c}\,\underline{f}_{ij}+m_{i}\,\underline{g}
\label{eq:04}\\
J_{i}\dfrac{d\underline{\omega}_{i}}{dt} 
& =\sum\limits _{\left\{ i,j\right\} =c}\,\underline{l}_{ij}\times\underline{f}_{ij}
\label{eq:05}
\end{align}
with mass $m_{i}$, position $\underline{r}_{i}$, moment of inertia
$J_{i}$, and angular velocity $\underline{\omega}_{i}$ of particle
$i$. The right hand side terms in Eq.\ (\ref{eq:04}) are the sum of the inter-particle
forces $\underline{f}_{ij}$ due to contacts $c$ with particles $j$
and volume/body forces, here from gravity $\underline{g}$.
The right hand side in Eq.\ (\ref{eq:05}) is the torque arising from the contacts
$c$ with the branch vector $\underline{l}_{ij}$, i.e., the distance
vector from the particle center to the contact point of the two particles
$i$ and $j$.

The inter-particle forces $\underline{f}_{ij}$ are modeled by well-known
force-overlap relations in combination with capillary forces, which
are induced by liquid bridges between the interacting particles. Details
of the contact force modeling are given in the next subsection.

\subsection{Contact force law}

\begin{center}
\begin{figure}[h]
\centering
\includegraphics[width=0.4\textwidth]{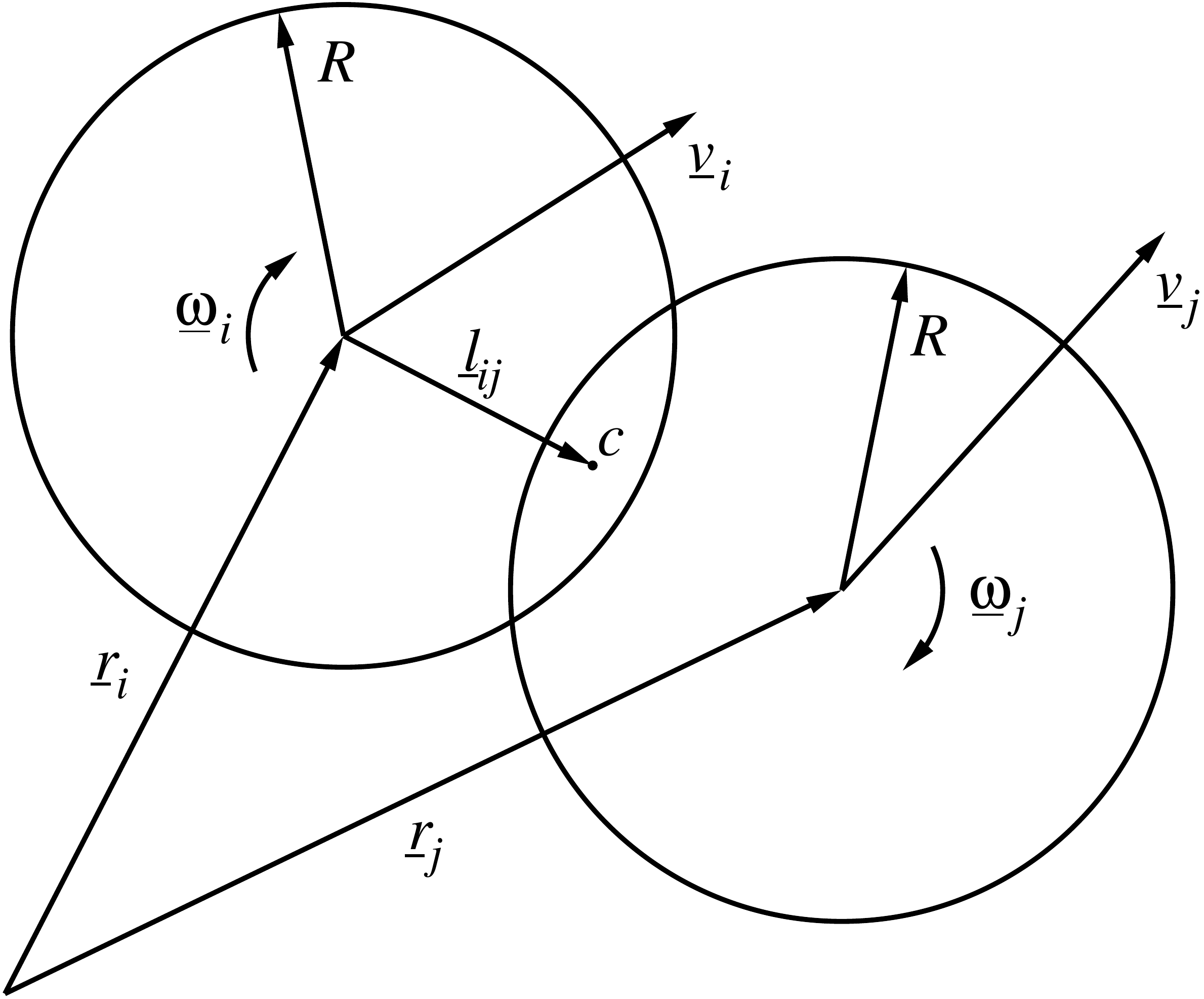}
\caption{Sketch of the contact between two particles $i$, $j$.}
\label{fig:04}
\end{figure}
\end{center}

Since the realistic modeling of the deformation of two interacting
particles, e.g.\ in a core shooting material, is much too complicated, 
the inter-particle force is described by a force-overlap relation. The
modeling of the contact force is based on the quantities/symbols given in 
Figs.\ \ref{fig:04} and \ref{fig:05}.

The force $\underline{f}_{ij}$ on particle $i$, from particle $j$, at contact $c$, 
can be decomposed into a normal and a tangential part as 
\begin{align}
\underline{f}_{ij} & = f_{ij}^{n}\,\hat{n}_{ij} + f_{ij}^{t}\,\hat{t}_{ij} + f_{ij,c}\,\hat{n}_{ij} \,.
\label{eq:06}
\end{align}
with normal $\hat{n}_{ij}$ and tangential unit vector $\hat{t}_{ij}$ at the contact point $c$.
The normal force $f_{ij}^{n}$, the tangential force $f_{ij}^{t}$ 
and the capillary force $f_{ij,c}$ due to a liquid bridge between particle $i$ and $j$ are specified below. 

\noindent 

Two dry particles $i$, $j$ with radius $R$, which are moving with
velocities $\underline{v}_{i}$ and $\underline{v}_{j}$ and rotating 
with angular velocities $\underline{\omega}_{i}$ and $\underline{\omega}_{j}$, interact,
if the normal overlap $\delta$ is positive
\begin{align}
\delta & =2R-\left(\underline{r}_{i}-\underline{r}_{j}\right)\cdot\hat{n}_{ij}>0 \, .
\label{eq:07}
\end{align}
The {\em normal contact force} involves a linear repulsive and a linear dissipative force, 
\begin{align}
f_{ij}^{n} & = k_{n}\delta+\gamma_{n}\, v_{ij}^{n}\,,
\label{eq:08}
\end{align}
with normal spring stiffness $k_{n}$, normal viscous damping $\gamma_{n}$ 
and normal velocity $v_{ij}^{n}$.

Our model is able to capture
friction forces and torques, as well as rolling and torsion torques,
as described in Ref.\ \cite{Lud08GM}. For the sake of brevity
we did set the latter interactions to zero and focus on friction only,
i.e.\ the tangential friction force is 
\begin{align}
f_{ij}^{t} & = k_{t}\chi+\gamma_{t}\, v_{ij}^{t}\,,
\label{eq:09}
\end{align}
with tangential spring stiffness $k_{t}$, tangential viscous damping $\gamma_{t}$ 
and tangential velocity $v_{ij}^{t}$, where $\chi$ is the integral of $v_{ij}^{t}$
over time, adapted such that the tangential force is limited by Coulomb sliding friction 
\begin{align}
f_{ij}^{t} & \leq \mu_C\, f_{ij}^{n}\,,
\label{eq:10}
\end{align}
with Coulomb's coefficient of friction $\mu_{C}$.

Note, that ${\displaystyle f_{ij}^{n}}$ and $f_{ij}^{t}$ give only non-zero contributions to 
$\underline{f}_{ij}$, when the two particles are in contact, $\delta>0$. The {\em capillary force} 
$f_{ij,c}$ is also active when two particles separate after a contact. Details of the 
modeling of $f_{ij,c}$ are given in the next subsection.

\subsection{Capillary forces}
\label{sec:capmodel}
\begin{center}
\begin{figure}[h]
\centering
\includegraphics[width=0.2\textwidth]{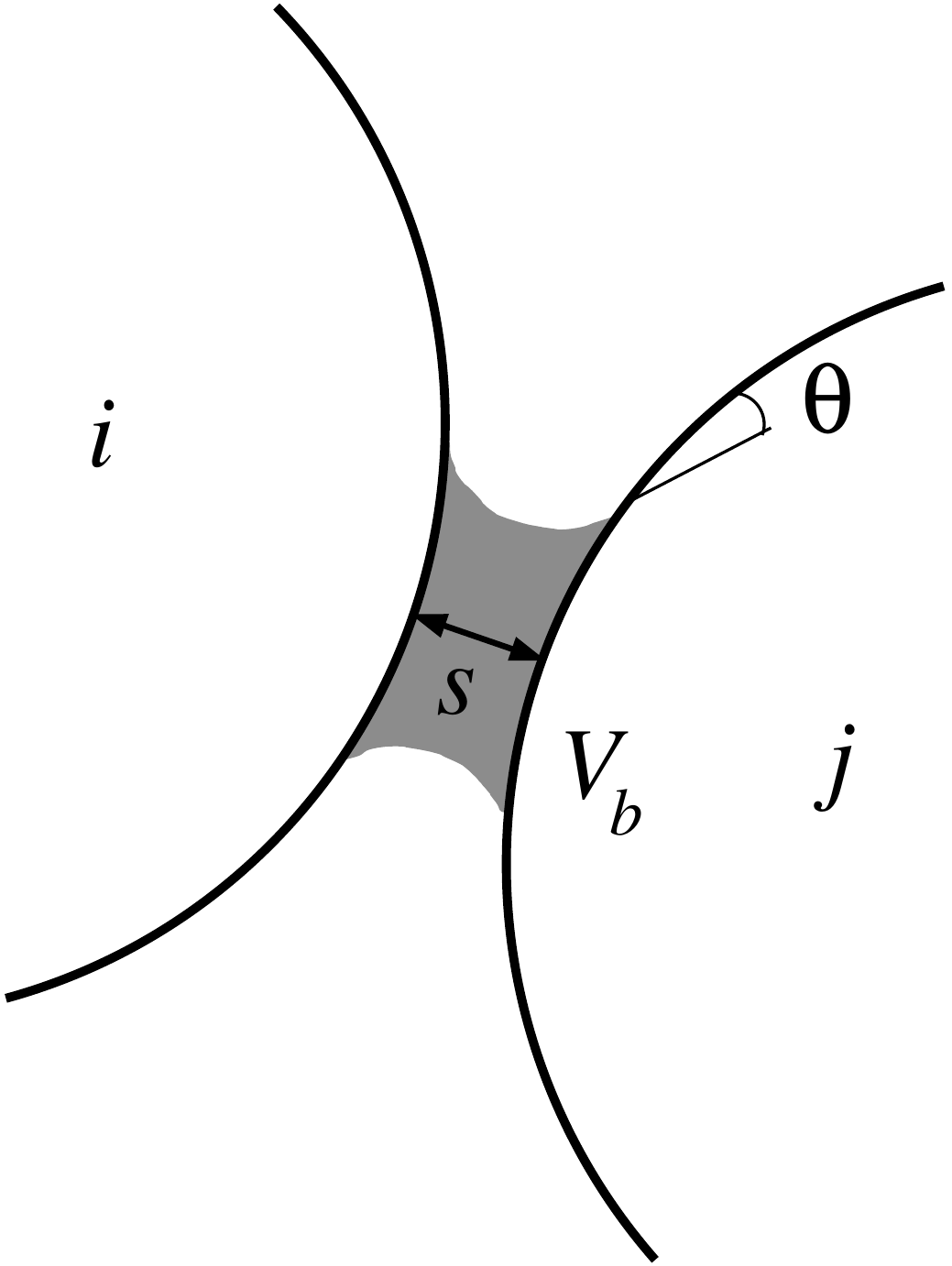}
\caption{Sketch of a liquid bridge between particles $i$ and $j$ 
with the bridge length $s$, the bridge volume $V_{b}$ and the contact angle $\theta$.
\label{fig:05} }
\end{figure}
\end{center}

The shapes of liquid bridges between individual particles of a granular
medium depend strongly on the amount of the added liquid, see 
e.g.\ Refs.\ \cite{Wei99,Her05}.
For core shooting materials with low mass ratios between binder and 
dry sand or sand-like materials
we expect, based on the findings in \cite{Her05}, that the grains
are connected by individual capillary bridges. The relevant parameters
of such a bridge are indicated in Fig.\  \ref{fig:05}.

\noindent 

The liquid volume of the bridge is $V_{b}$. The length $s=-\delta$ 
of the bridge is given by the surface distance of the two particles,
which are connected by the bridge. Finally, the equilibrium contact angle
$\theta<90^{o}$ is found at the bridge-particle contact line.

With these parameters, we approximate the inter-particle force
$f_{ij,c}$ of the capillary bridge according to the proposal of Willett et al. 
\cite{Wil00}, see also \cite{Her05,Rad10}, who calculated $f_{ij,c}$ with the gorge
method at the bridge neck
\begin{align}
f_{ij,c} & = \frac{2\,\pi\,\gamma\, R\, \cos\left(\theta\right)}{1 + 1.05 \hat{s} + 2.5 \hat{s}^2} \,.
\label{eq:11}
\end{align}
with surface tension, $\gamma$ and dimensionless $\hat{s}=s\,\sqrt{R/V_b}$.
Note, that $f_{ij,c}$ exists only during and past a contact between particle
$i$ and $j$, providing a non-zero contribution to $\underline{f}_{ij}$ 
until the total distance $s$ between $i$ and $j$ rises above the critical bridge length 
$s>s_{crit}$. Then, the bridge ruptures and  $f_{ij,c}$ becomes zero.
Several authors have proposed correlations between $s_{crit}$ and
other parameters of the capillary bridge \cite{Lia93,Wei99}. We use
the approximation of Willett et al.\ \cite{Wil00}
\begin{align}
s_{crit} & = R \,\left(1 + \frac{1}{2}\,\theta\right)\,\left[\left(\frac{V_{b}}{R^3}\right)^\frac{1}{3}+ 0.1\,\left(\frac{V_{b}}{R^3}\right)^\frac{2}{3}\right] \,.
\label{eq:12}
\end{align}

The model equations presented above were implemented into
the open-source DEM software package LIGGGHTS, version 2.3.2.
The simulations were carried out at the HPC cluster CVC at the
University Computer Center of the TU Bergakademie Freiberg.

\subsection{Capillary force model validation}
\label{sec:LBval}

\begin{figure}
\begin{centering}
\includegraphics[width=0.4\textwidth]{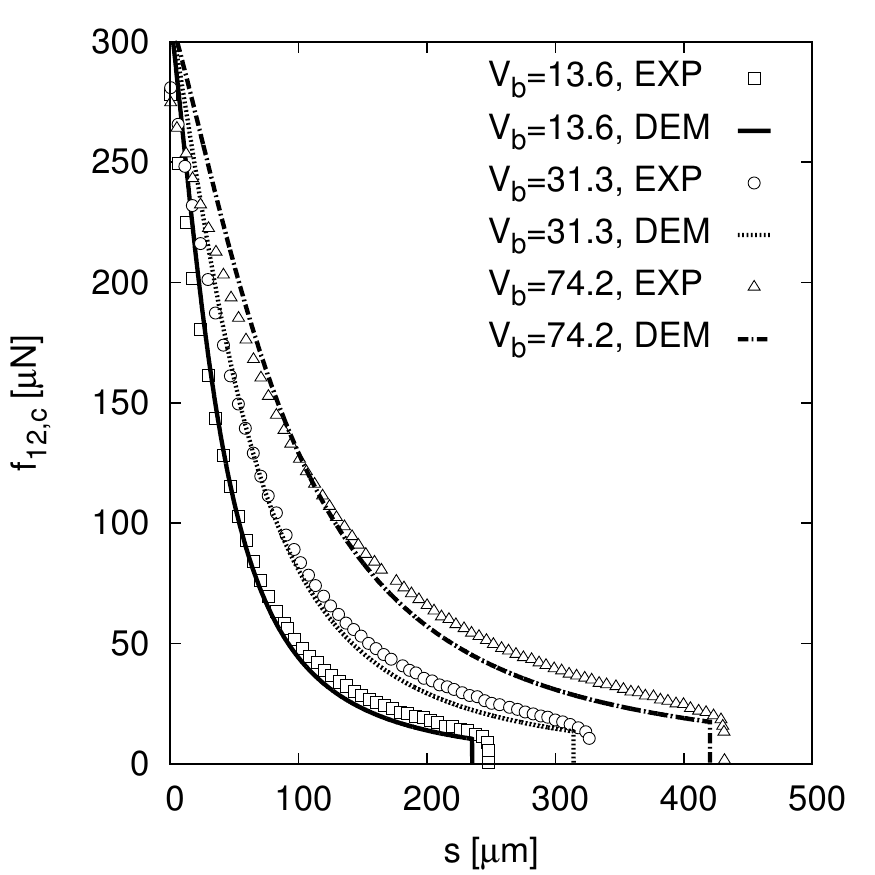}
\end{centering}
\caption{
Forces in a capillary bridge with contact angle $\theta=0^{\circ}$
between two equal-sized particles; lines indicate results of the DEM simulation, points
give the experimental data of Willett et al.\ \cite{Wil00}. Liquid bridge volumes $V_b$ 
are given in nano-liters ($10^{-9}$l$=$nl).
\label{fig:07}}
\end{figure}

For the validation of the capillary force calculation, we perform DEM simulations 
according to the experiments described in Willett et al.\ \cite{Wil00}. There, the separation 
process of two equal-sized particles with $R=2.4$\,mm connected by a capillary bridge has been 
investigated. The liquid surface tension was $\gamma=20.6$\,mN/m. The particles were perfectly 
wetted by the liquid, i.e.\  $\theta=0^{\circ}$. Experiments with  different liquid bridge 
volumes $V_{b}=13.6,\, 31.3$ and $74.2$\,nl have been carried out. 

In the DEM simulations, we track the capillary force $f_{12,c}$ due to Eq.\ (\ref{eq:11}) when the
particles separate with velocity $v_{12}^{n}=0.001\,\mbox{m/s}$ from
the distance $s=0$ (at the end of the mechanical contact between the two particles)
until $s=s_{crit}$ (the rupture distance, Eq.\ (\ref{eq:12})).

Fig.\ \ref{fig:07} gives a comparison of the DEM results and the experimental data of Willett 
et al.\ \cite{Wil00}, showing the good agreement between our data and the measurements.

\subsection{Setup of the numerical rheometer}

\begin{figure}[h]
\noindent \begin{centering}
\includegraphics[width=0.465\textwidth]{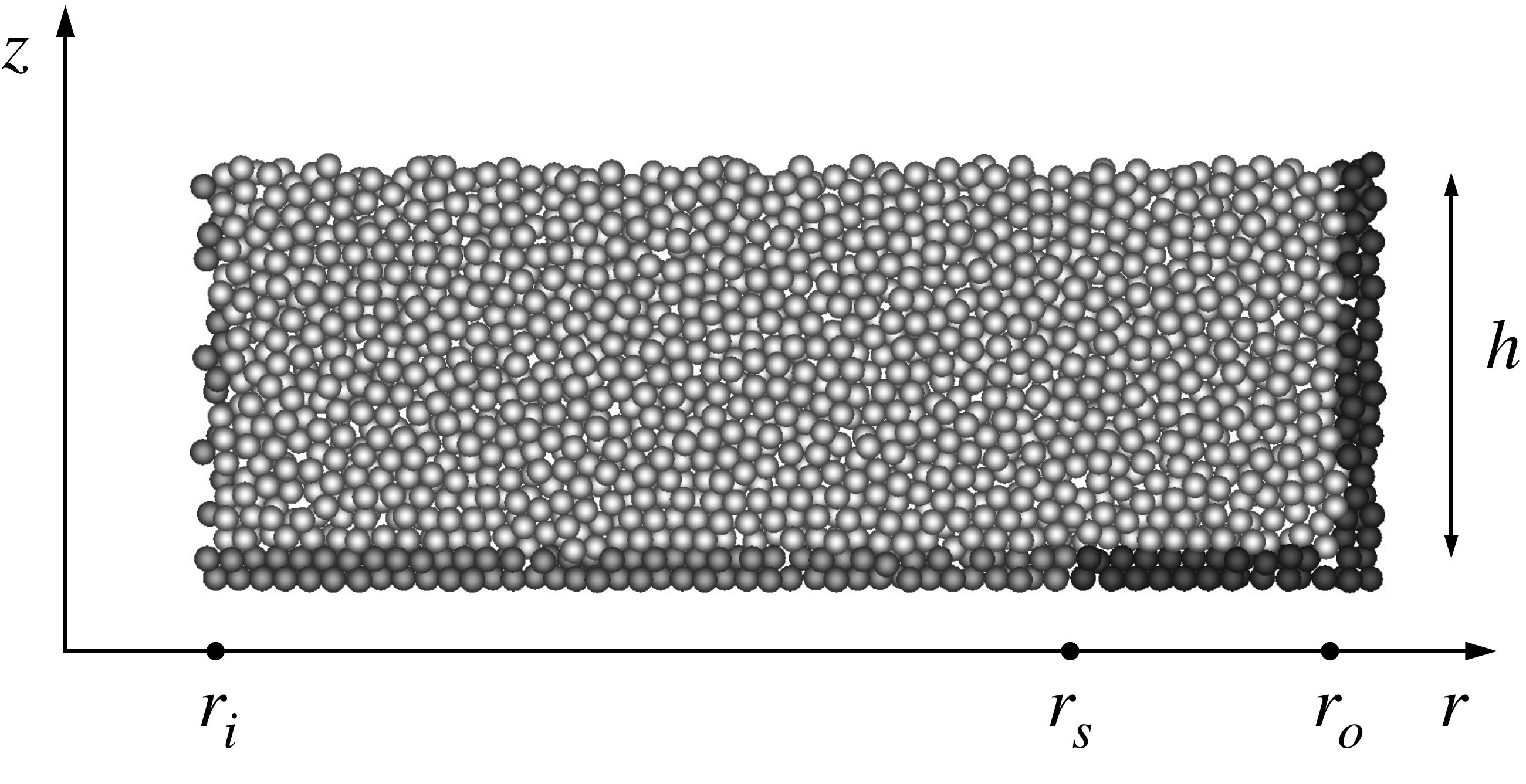}
\par\end{centering}
\caption{
Setup of the numerical rheometer, light gray: sheared granular material,
medium gray: static, inner part of the shear cell, dark gray: rotating, outer 
part of the shear cell (the medium and dark gray particles are part of
the rough walls of the shear cell and thus displayed as particles).
}
\label{fig:06}
\end{figure}

The flow and rheology of dry and wet granular materials in a three-dimensional
split bottom shear cell \cite{Fen04,Lud08,Lud08Pa,Lud11} are investigated with
the DEM model. Basic parameters of the shear cell geometry are sketched in 
Fig.\ \ref{fig:06}, with the values: 
$r_{i}=14.7$\,mm, $r_{s}=85$\,mm, $r_{o}=110$\,mm, and $U_{o}=6.9$\,mm/s. 

In contrast to previous studies, where only a quarter of the rheometer was modeled,
see Refs.\ \cite{Lud08,Lud08Pa,Lud11}, here we study the full ring-geometry.
The rheometer is filled with $n_{P}\simeq210000$ particles, which
all have the same diameter $d_{P}=2$\,mm, with density $\rho_{P}=2000$\,kg/m$^{3}$.
We assume, that polydispersity has in a first approximation little influence on 
the macroscopic flow behavior of the granular material, similar to the findings 
in Refs.\ \cite{Gon10,Sha12}. The expected small effects of polydispersity should be 
investigated in more detail in the future, but are not subject of the present study.

The mechanical parameters of the particles are chosen as $k_{n}=110$\,N/m,
$\gamma_{n}=2\cdot10^{-3}$\,kg/s, $k_{t}=12$\,N/m, $\gamma_{t}=0.5\cdot10^{-3}$\,kg/s,
and $\mu_C=0.01$ (if not mentioned otherwise) in order to match those of the 
numerical simulations by Luding \cite{Lud08,Lud11} for straightforward validation 
purposes. 

An adaption of the contact model to real sand F35 is presently not possible and must 
be postponed since the material parameters are actually unknown. However, 
this is no essential constriction, because details of the contact model, including stiffness, 
have been found to cause only small differences for granular flows in the collisional and dense
regimes, as long as the particles are not too soft, see e.g.\ Refs.\ \cite{Lud94,Ots10}, where
the effect of softness of the material was studied in more detail.

The results of our calculations with dry granular material
($f_{c}=0$) are used to verify and validate the DEM code and
force model implementation, in comparison with previous 
results \cite{Lud08}.  The simulations with wet granular
materials should reveal if the influence of the liquid bridges
(the model of which was validated in subsection \ref{sec:LBval})
on the apparent shear viscosity is the same as in the experiments. 

For the simulations of wet granular materials, the equilibrium contact angle 
$\theta$ was varied between $0^{\circ}$ and $20^{\circ}$, whereas the
surface tension $\gamma = 20.6\,$mN/m was kept constant. The simulations
were conducted for two different bridge volumes $V_b$ ($4.2$\,nl, $42$\,nl) 
for each capillary bridge. Assuming 
an average number $n_{c,i} = 6$ of capillary bridges per particle,
this corresponds approximately to a mass ratio $m_{l}/m_{ap}$ = 0.15\%, 1,5\%
between the liquid and the dry granular material, respectively. Here, the total mass 
$m_{l}=\sum_{i=1}^{n_P} \rho_l\,V_b\, n_{c,i}/2$ of the liquid is calculated with an 
arbitrary fixed value $\rho_l$ = 1000 kg/m$^3$, because $\rho_l$ is not 
applied in the DEM model. The total mass of all dry particles
is $m_{ap} = \sum_{i=1}^{n_P} m_i$. Note, that $m_{l}/m_{ap}$  was found to 
change during the simulation, because $n_{c,i}$ varies in the interval 
$6 \lesssim n_{c,i} \lesssim 7$, however this small effect was ignored
and $n_{c,i}=6$ was kept constant for the calculation of $V_b$.

The influence of other, more complex liquid distributions 
is analyzed in ongoing simulations and will be reported elsewhere.

\section{Results}

\subsection{Micro-macro transition}

As in Refs.\ \cite{Lud08,Lud08Pa,Lud11}, 
continuum quantities like the solid-fraction $\phi$, the velocity-field 
$\underline{u}$ or the stress-field $\underline{\underline{\sigma}}$ are computed
by a micro-macro transition method from the DEM results, e.g.,
\begin{align}
\phi(\underline{r}) &= \frac{1}{\Delta t\,\Delta V}\int\limits _{\Delta t}\
\sum_{i\in\Delta V}V_{i} dt \,,
\label{eq:13}\\
\underline{u}\left(\underline{r}\right) 
& =\frac{1}{\Delta t\,\Delta V}\int\limits _{\Delta t}\left(\sum_{i\in\Delta V}V_{i}\,\underline{v}_{i}\right)dt  \times \frac{1}{\phi(\underline{r})} \,,
\label{eq:14}\\
\underline{\underline{\sigma}}\left(\underline{r}\right) 
& =\frac{1}{\Delta t\,\Delta V}\int\limits _{\Delta t}\left(\sum_{i\in\Delta V}m_{i}\underline{v}'_{i}
\otimes\underline{v}'_{i}+\sum_{c\in \Delta V}^{\left\{ i,j\right\} =c}\underline{f}_{ij}
\otimes\underline{l}_{ij}\right)dt\,,
\label{eq:15}
\end{align}
with fluctuation velocity $\underline{v}'_{i}=\underline{v}_{i}-\underline{u}(\underline{r})$,
averaging time intervals of typically $\Delta t=5$\,s, 
a discrete averaging time-step $dt=0.05$\,s, and particle volume $V_i$, 
together with the (ring/torus-shaped) averaging volume $\Delta V$ at various 
positions $\underline{r}=(r,z)$ in the system (with cylindrical coordinates).
Further parameter-fields like strain rate magnitude, shear stress magnitude, 
hydrostatic pressure, apparent viscosity and inertial number 
can be calculated from these variables as:
\begin{align}
\dot{\gamma} 
& =\frac{1}{2} \sqrt{\left(\frac{\partial u_{\varphi}}{\partial r} -\frac{u_{\varphi}}{ r}\right)^2 
    + \left(\frac{\partial u_{\varphi}}{\partial z}\right)^2} \,,
\label{eq:16} \\
\left|\tau\right| 
& =\sqrt{\sigma_{r\varphi}^{2}+\sigma_{z\varphi}^{2}} \,,
\label{eq:17} \\
p 
& =\frac{1}{3} \left(\sigma_{rr} + \sigma_{zz}  + \sigma_{\varphi\varphi}\right) \,,
\label{eq:18} \\
\eta
& = \frac{\left|\tau\right|}{\dot{\gamma}} \,,
\label{eq:19} \\
I
& = \frac{\dot{\gamma}\, d_{P}}{\sqrt{{p}/{\rho}}} \,.
\label{eq:20}
\end{align}
In contrast to the experimental setup, these parameters can be investigated
\emph{locally}, i.e.,\ at arbitrary positions $\underline{r}$ anywhere in 
the filled measurement volume (gap) of the rheometer.

All simulations which are discussed afterwards run for 20 s. For the 
average, only the period between 15 s and 20 s are is into account. 
Therefore, the system is examined in quasi-steady state flow conditions
(the transient behavior at the onset of shear is disregarded).
However, it cannot be excluded, that long-time relaxation 
effects may have an impact on our findings, which
is not adequately resolved by our relatively short simulations.

\subsection{Dry granular material}

\begin{figure}[h]
\noindent \begin{centering}
\includegraphics[width=0.45\textwidth]{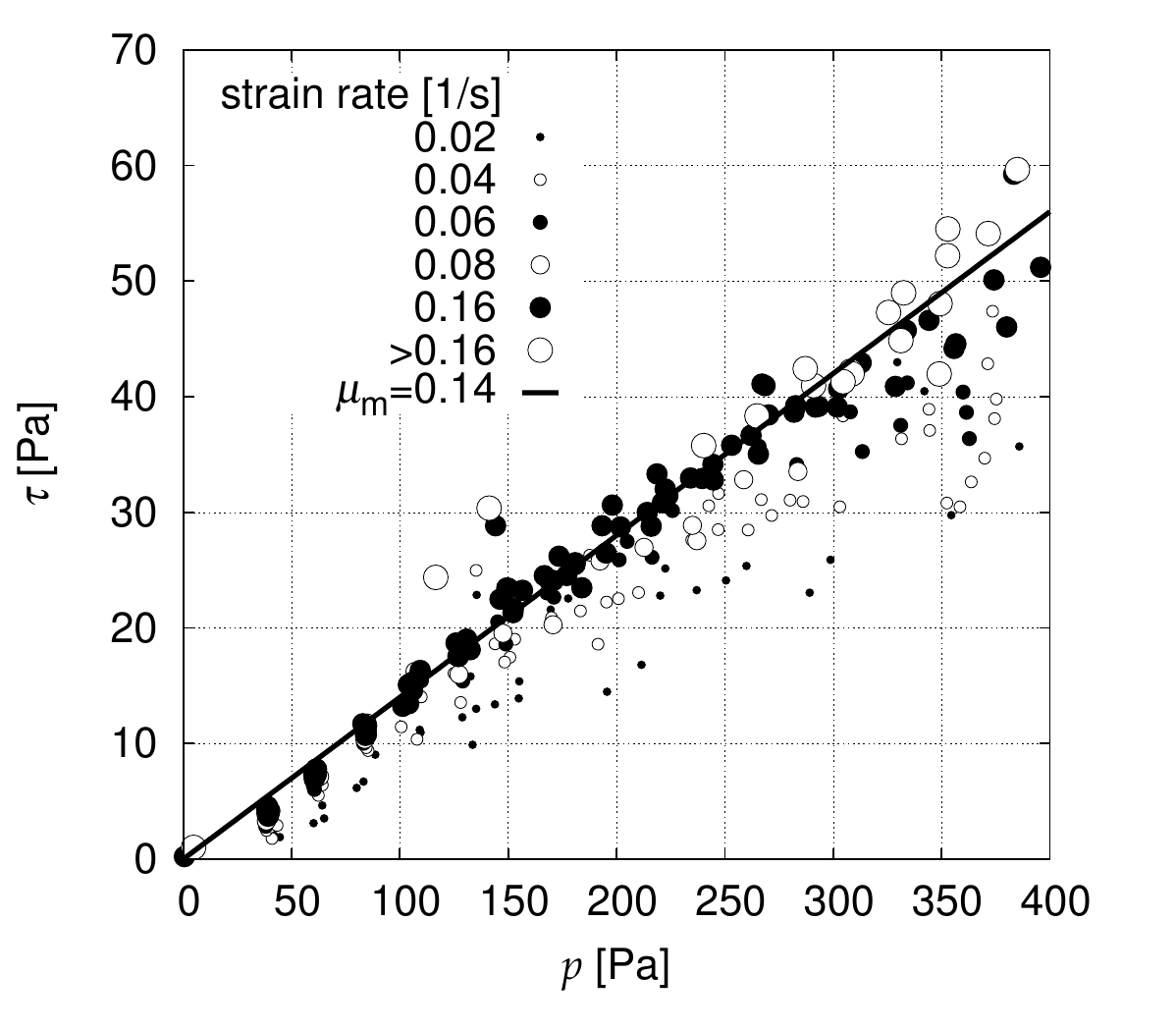}
\par\end{centering}

\caption{
Local shear stress $\left|\tau\right|$ plotted against local 
pressure $p$.  Different symbol sizes indicate the magnitude of the strain 
rate $\dot{\gamma}$ at different locations within the rheometer gap, 
with $\dot{\gamma}$ given in the inset in units of $\mbox{s}^{-1}$;
larger symbols correspond also to larger $\dot{\gamma}$.
The solid line represents the function $\left|\tau\right|=\mu_m\, p$,
with the macroscopic friction coefficient $\mu_m=0.14$.
\label{fig:08}}
\end{figure}

Fig.\ \ref{fig:08} compares the shear stress intensity $|\tau|/p = \mu_m$
in our DEM simulations to the previous findings of Luding \cite{Lud08,Lud11},
with the same macroscopic friction value $\mu_m = 0.14$ for $\mu_C=0.01$ as
reported earlier.

\begin{figure}[h]
\noindent \begin{centering}
\includegraphics[width=0.45\textwidth]{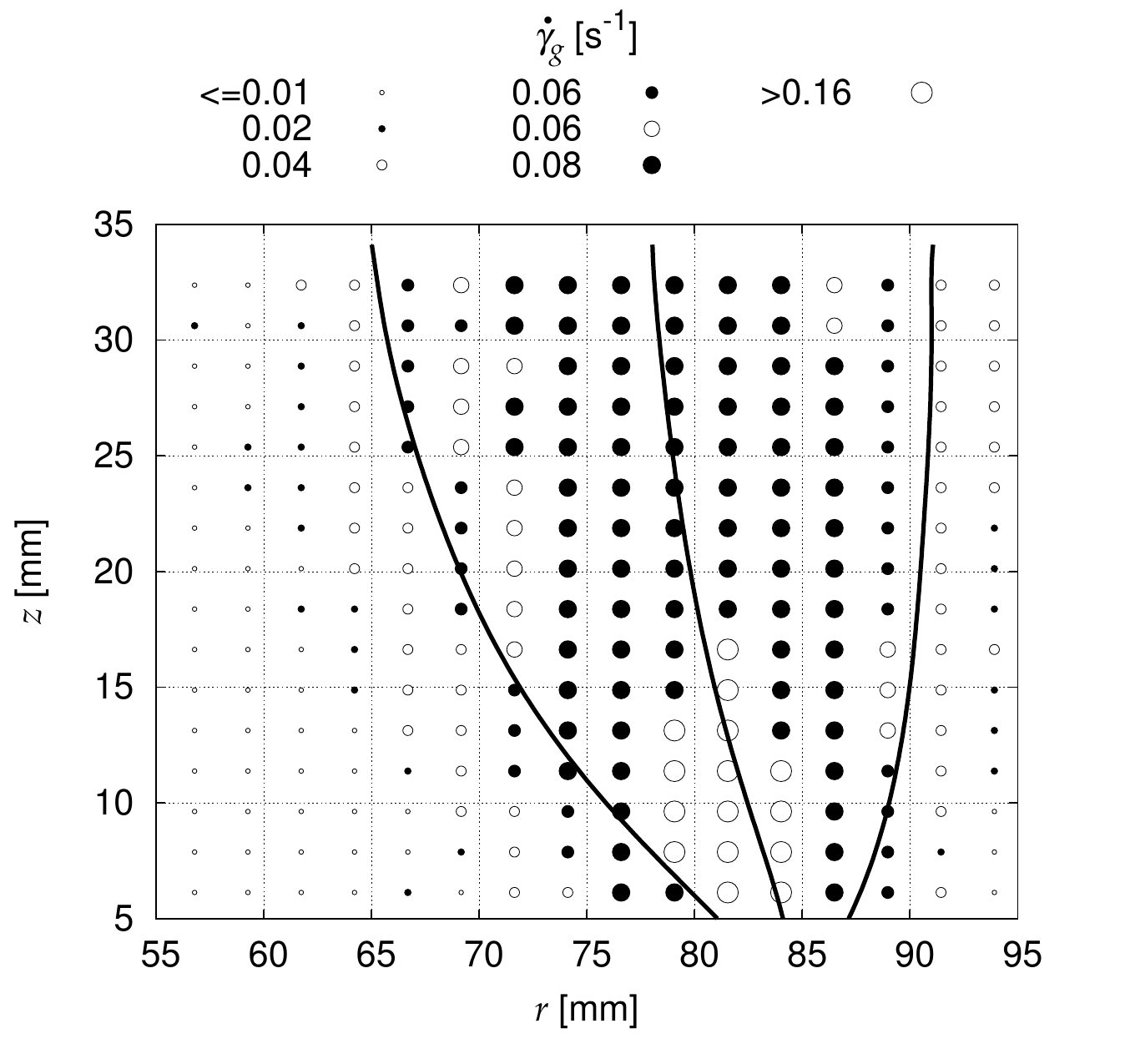}
\par\end{centering}

\caption{
Strain rate $\dot{\gamma}$ as function of radial and vertical position 
within the rheometer gap. As in Fig.\ \ref{fig:08}, different symbol
sizes indicate the magnitude of $\dot{\gamma}$. The lines 
indicate the center location $R_c$ (middle line) and the width $W$ of the shear
bands (outer lines), as obtained from the fit function Eq.\ (\ref{eq:19}).
\label{fig:09}}
\end{figure}

In order to allow for a more quantitative analysis of the dynamic 
behavior of the dry granular material, its shear band structure is analyzed.
The shear bands in dry granular matter
have been intensively studied in experiments and DEM simulations,
see e.g., Refs.\ \cite{Fen04,Lud08}. Among other details, as will be 
discussed elsewhere, the profiles of the velocity
field are well approximated by error functions:
\begin{align}
\omega \left(r\right) = A + B\, \mathrm{erf}\,\left(\frac{r-R_c}{W} \right)
\label{eq:21}
\end{align}
where the dimensionless amplitudes are $A \simeq B \simeq 0.50$, $R_c$ 
is the center of the shear band, and $W$ is its width. The strain rate as function of 
$r$ and $z$ and the shear band structure in the rheometer gap are shown in Fig.\ \ref{fig:09}.

Obviously, our results for both the shear stress intensity and the shear band structure 
agree well with the findings in \cite{Lud08}, which were obtained with other software
for DEM simulation and analysis. Therefore we conclude, 
that there are no implementation errors in the basic LIGGGHTS code and our 
dry contact model implementation, for the parameters used here.

\subsection{Wet granular material}

\subsubsection{Macroscopic friction and cohesion}
\begin{figure}[h]
\noindent \begin{centering}
\includegraphics[width=0.45\textwidth]{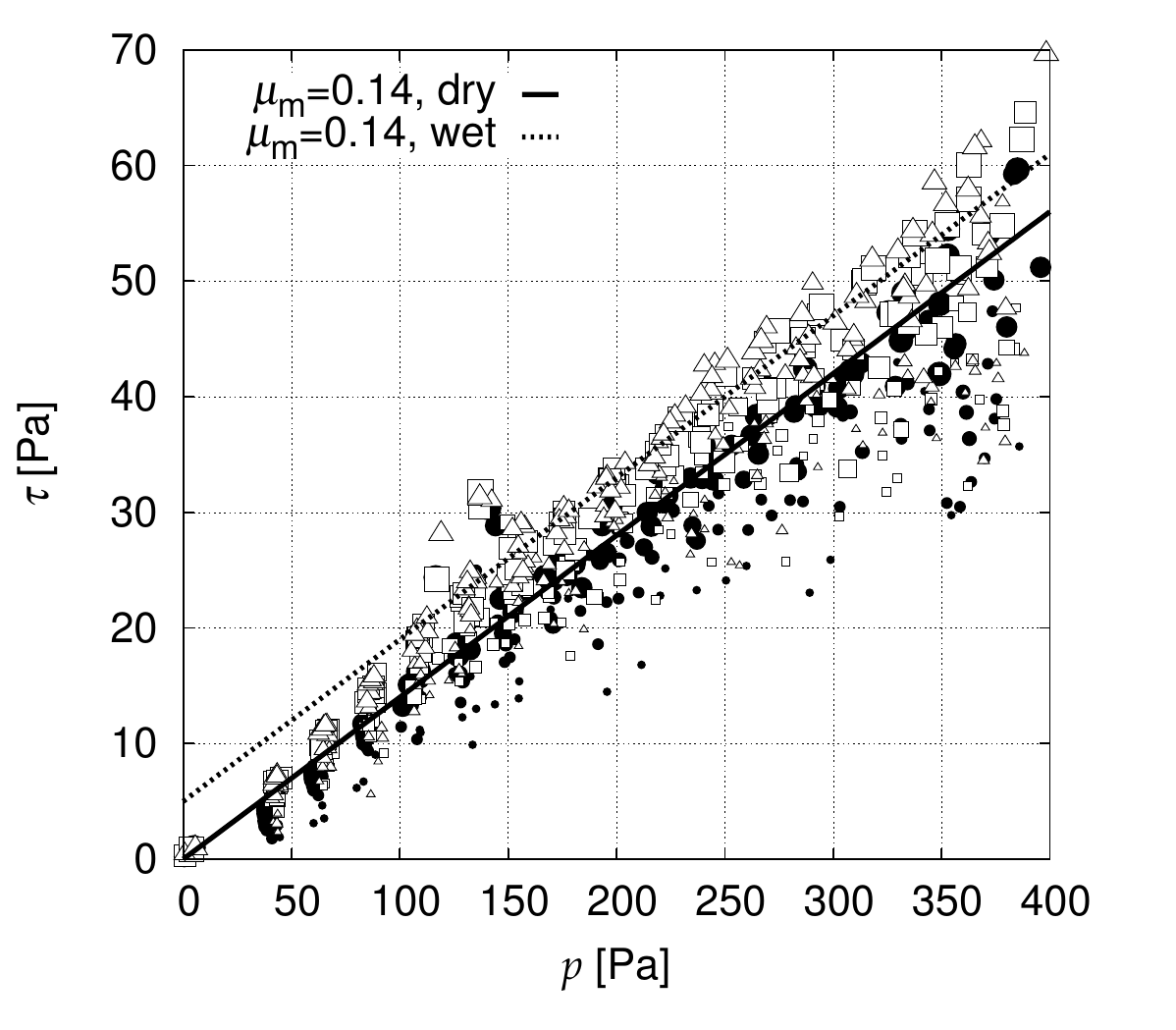}
\par\end{centering}

\caption{
Shear stress $\left|\tau\right|$ plotted against pressure $p$.
Different symbols indicate results from dry material ($\bullet$), 
and from two wetted materials, which contain liquid bridges with
(i) $\theta = 0^{\circ}$, $V_b = 4.2$\,nl ($\Box$) and
(ii) $\theta = 20^{\circ}$, $V_b = 42$\,nl ($\triangle$).
The magnitude of the strain rate $\dot{\gamma}$  is indicated by the 
size of the symbols similar as in Fig.\ \ref{fig:08}.
The solid and the dashed lines represent the function 
$\left|\tau\right|=\mu_m\,p + c$,
with the macroscopic friction coefficients $\mu_m$ for the
dry ($c=0$) and the wet ($c=5$ Pa) materials, respectively.
}
\label{fig:10}
\end{figure}

In Fig.\ \ref{fig:10}, the dry results are compared to two simulations 
with liquid bridges. The addition of the
liquid bridge forces leads to larger shear stress magnitudes
so that the macroscopic yield stress at critical state 
(the termination locus, reached after large shear strain)
is shifted upwards.
The offset on the vertical axis is
referred to as the macroscopic cohesion $c$.
While the dry data are fitted well by the macroscopic line
fit with $c=0$, the wet data for small $p<100$\,Pa considerably
drop below the fit result with constant $c>0$.

\subsubsection{Local shear viscosities}
\begin{figure}
\begin{centering}
\includegraphics[width=0.49\textwidth]{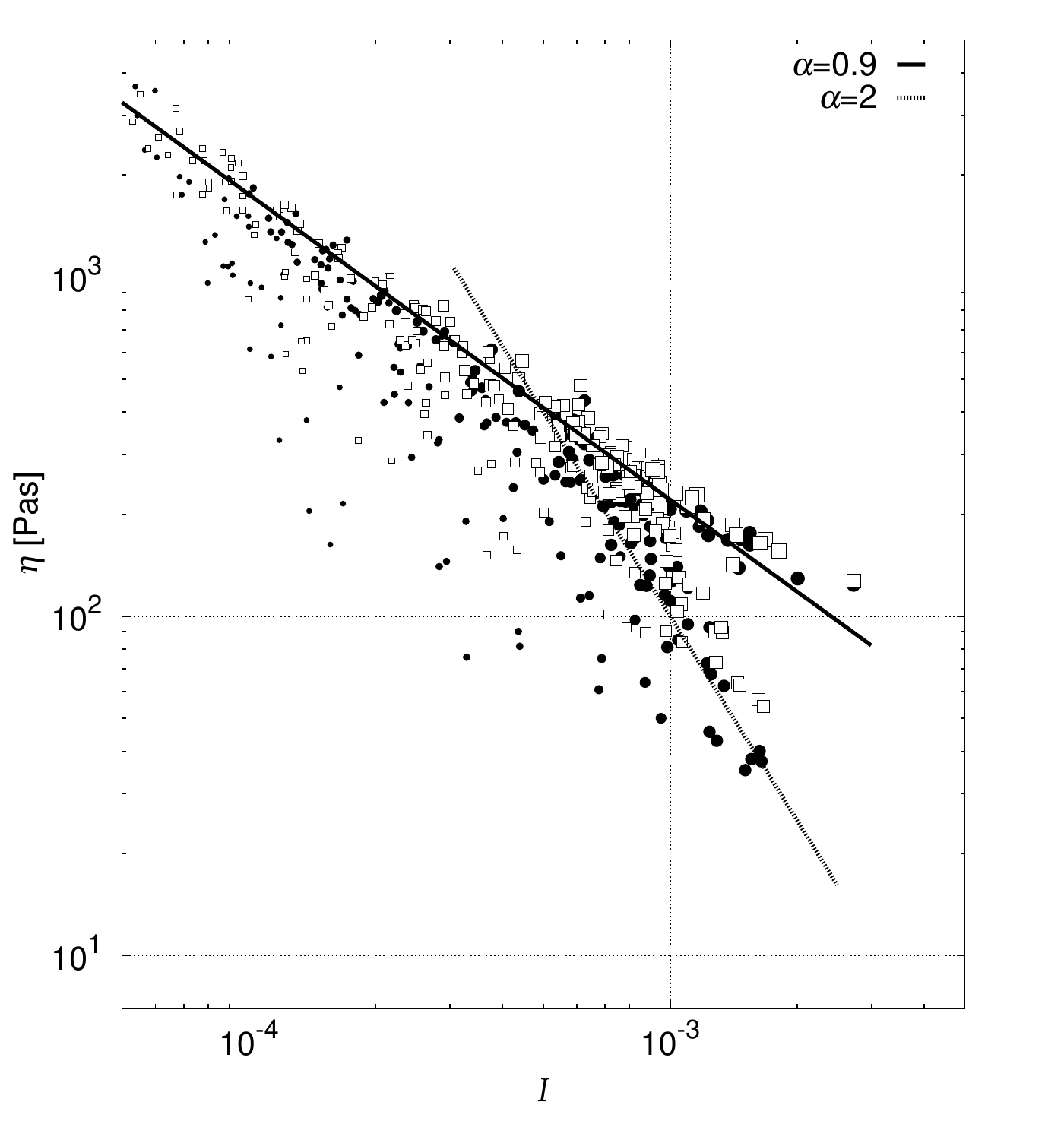}\\
\includegraphics[width=0.49\textwidth]{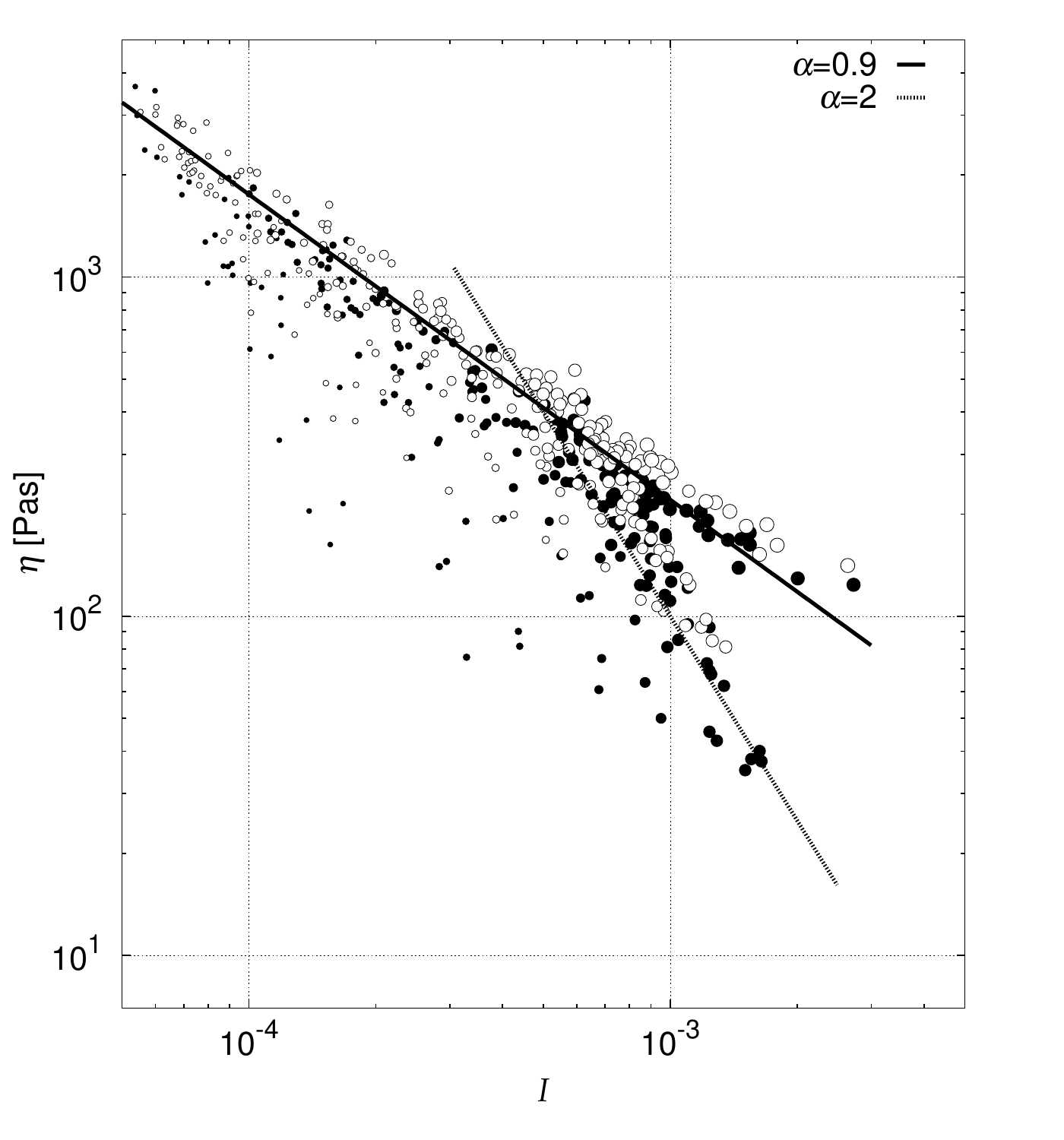}
\par\end{centering}
\caption{
Local shear viscosity $\eta\left(I\right)$ from the DEM simulations.
Different symbols indicate results from dry ($\bullet$) and wet material 
(open symbols). The wet material contains liquid bridges 
with (top) $V_b = 4.2$\,nl ($\Box$) and (bottom) $V_b = 42$\,nl ($\circ$), 
whereas $\theta = 0^{\circ}$ is the same for both cases.
The magnitude of the local strain rate $\dot{\gamma}$  is indicated by
the size of the symbols similar as in Fig.\ \ref{fig:08}.
\label{fig:11}}
\end{figure}

\begin{figure}
\begin{centering}
\includegraphics[width=0.49\textwidth]{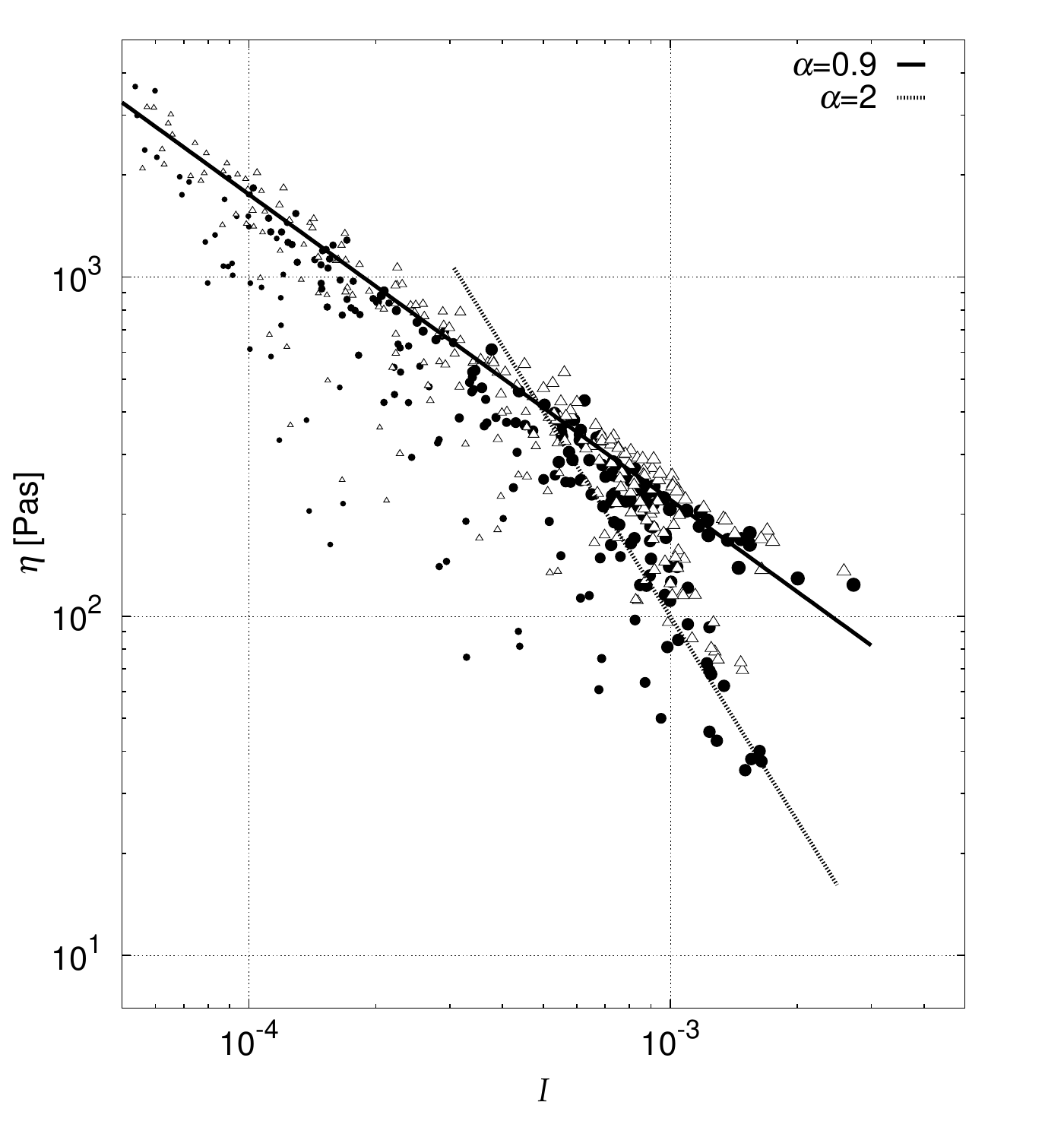}\\
\includegraphics[width=0.49\textwidth]{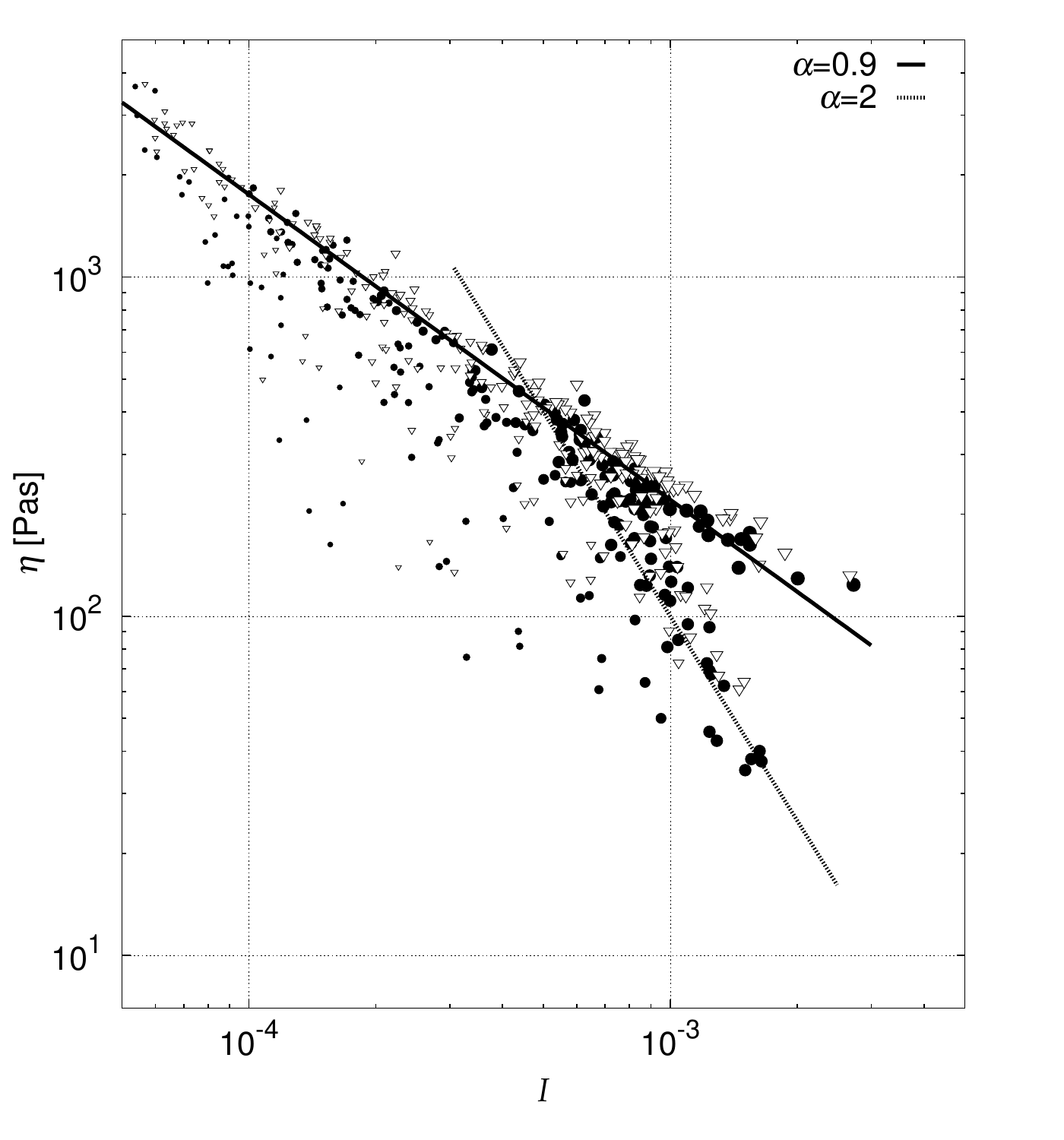}
\par\end{centering}

\caption{
Apparent shear viscosity $\eta\left(I\right)$ from the DEM simulations.
Different symbols indicate results from dry ($\bullet$) and wet material 
(open symbols). The wet material contains liquid bridges 
with (top) $\theta = 10^{\circ}$ ($\triangle$)
and (bottom) $\theta = 20^{\circ}$ ($\triangledown$), whereas $V_b = 42$\,nl 
is constant in all cases.
The magnitude of the strain rate $\dot{\gamma}$  is indicated by the 
size of the symbols similar as in Fig.\ \ref{fig:08}.
\label{fig:12}}
\end{figure}

Next, the local shear viscosities $\eta$ are compared for dry and different 
wet granular materials. Figs. \ref{fig:11} and \ref{fig:12} give the correlations
$\eta\left(I\right)$, which are found from several DEM simulations with different 
liquid bridge volumes $V_b=4.2$\,nl, $42$\,nl and contact angles $\theta = 0^{\circ}$,
$10^{\circ}$ and $20^{\circ}$.
The DEM simulations are evaluated locally in the granular medium.
Therefore, in contrast to the experiments, a single DEM simulation provides 
many data-points at various bulk-densities, shear-stresses, pressures and 
shear-rates that are plotted as different symbols for different $V_b$ and $\theta$.

As in the experiments, the inverse proportional dependence of the apparent viscosity 
on $I$ is evidenced for the dry and wet material, and, as to be expected,
the addition of small amounts of liquid increases the viscosity
by about 10 to 20\%. (The change appears not that large only due to the 
logarithmic vertical axis in the plots). The scaling $\eta \sim I^{-\alpha}$ 
changes from $\alpha=0.9$ to $\alpha=2$ in all materials,
in different regimes.
The influence of the contact angle and the liquid bridge volume $V_b$ on the 
rheology of the wet granular materials is presently investigated in more 
detail and results will be published elsewhere.

\subsubsection{Shear bands}

\begin{figure}[ht!]
\begin{centering}
\includegraphics[width=0.49\textwidth]{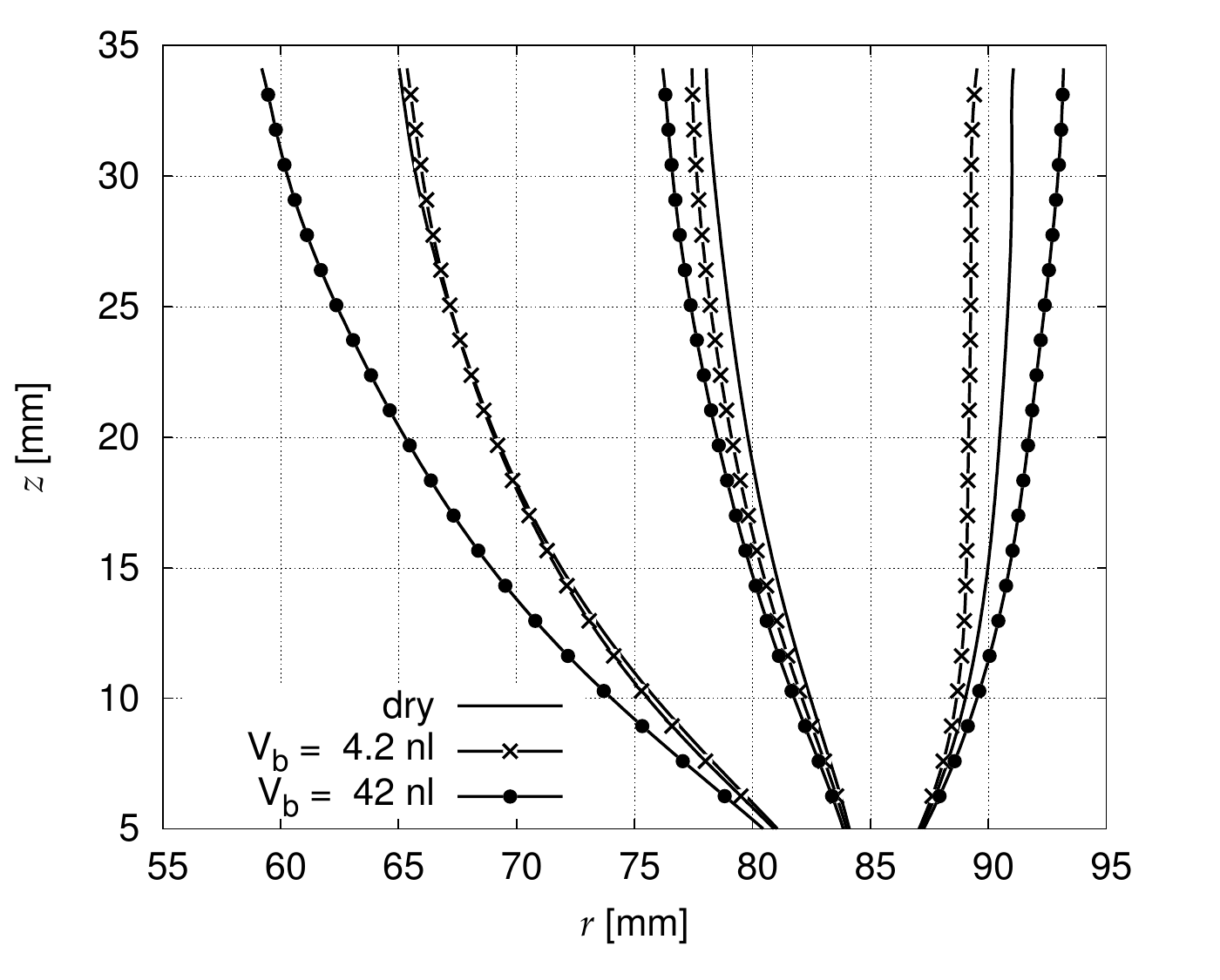}
\par\end{centering}

\begin{centering}
\includegraphics[width=0.49\textwidth]{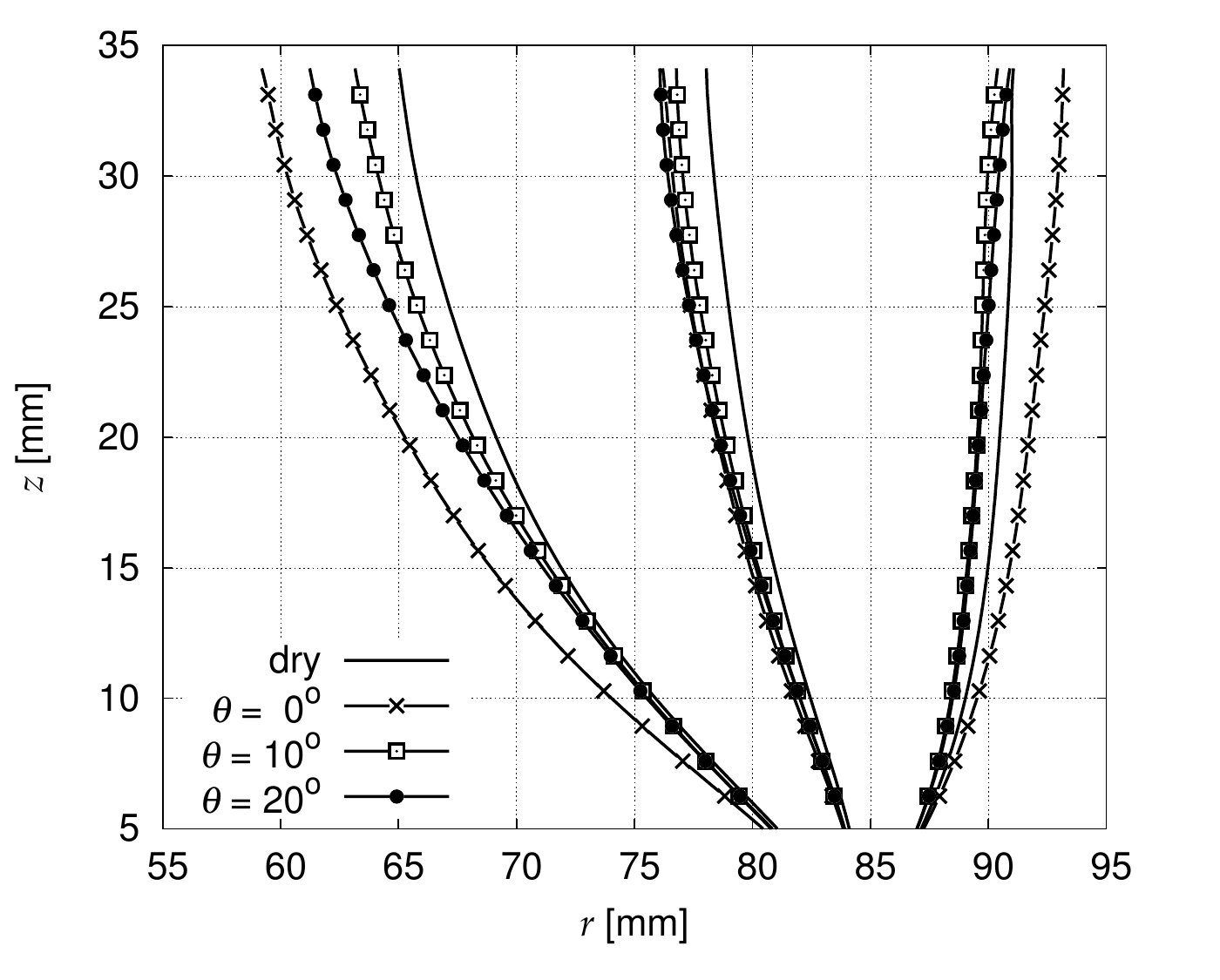}
\par\end{centering}

\caption{
Shear bands in dry and wet material, see Fig.\ \ref{fig:09}. 
Different line styles indicate results from dry (continuous) and wet material 
(dashed). Here, points serve only for distinction but don't represent flow data. 
In the upper subfigure, the wet material contains liquid bridges 
with (i) $V_b = 4.2$\,nl and (ii) $V_b = 42$\,nl, 
whereas $\theta = 0^{\circ}$ is constant in both cases.
In the lower subfigure, the wet material contains liquid bridges 
with (i) $\theta = 0^{\circ}$, (ii) $\theta = 10^{\circ}$ and (iii) $\theta = 20^{\circ}$, 
whereas $V_b = 42$\,nl is constant here.
\label{fig:13}}
\end{figure}

Next, the influence of the liquid on the dynamic behavior of the granular
material is investigated. The shear bands for dry and wet materials 
are given in Fig.\ \ref{fig:13}. Variations in $V_b$ and $\theta$ induce
noticeable changes in the shear band structure. With increasing liquid
content, its center position moves inwards to smaller radial distances. 
For small liquid content, the shear band width decreases, whereas it
increases for the larger liquid content simulations. 
The correlation for changing contact angle $\theta$ is more complex.
Due to the high liquid content, the shear band moves inwards. Not surprisingly, 
the shift is largest for lowest contact angle $\theta = 0^{\circ}$. But against 
expectation, the lowest shift is found for $\theta = 10^{\circ}$ and not for 
$\theta = 20^{\circ}$. Additionally, the qualitative change of the shear band 
as reported for very strong van der Waals type adhesion \cite{Lud11} is not 
reproduced here, possibly since the liquid bridge forces never become strong 
enough to have a similar effect. These interesting findings are presently 
investigated in more detail, results will be published elsewhere.

\section{Conclusions}

Shear experiments are complemented by a numerical rheometer study with core 
shooting materials as application in mind, but with a much more general perspective
concerning concepts and methods.  
The simple DEM contact and liquid bridge model
is validated using previous results from a three-dimensional 
split bottom ring shear cell for dry materials,
and with two-particle collision data from more advanced numerical 
and experimental studies for weakly wet materials. 

The DEM simulations of wet granular material show that the internal
structures of the sheared material, i.e., the shear bands, are qualitatively
the same as for dry materials. However, they move inwards with increasing liquid 
content and while getting a little narrower for small liquid content,
then become wider for very high liquid content. Finally, changing contact 
angles influence shear bands in non-trivial manner.

The apparent local shear viscosity of the granular material significantly increases 
when only small amounts of liquid are added to the material, representing well
the trend as seen from the experiments.

Future studies will involve a more quantitative study of the constitutive
relations that describe the rheology of the material and their implementation 
into CFD or FEM codes to predict large scale core shooting flow behavior.
The simulations and the experiments have to be performed in a more comparable
way between wet and dry configurations, i.e.\ using the same gap width;
the relation between the local and global viscosities (due to local and
global shear rates) has to be better understood.
An open question is how much a pendular liquid bridge contact 
model could be simplified, i.e.\ which details and non-linearities 
are important and which are not.
Nevertheless, more complex capillary bridge models have 
to be used, which allow the description of other than the 
pendular bridge regimes, too.

\section*{Acknowledgments}
We acknowledge the support of the ERASMUS program which allowed us to host
FU during his Master study at the University of Twente. Helpful discussions
with T. Weinhart, A. Singh, V. Magnanimo are appreciated. 
RS acknowledges the German Science Foundation (DFG) for funding parts 
of the work under project no. SCHW 1168/6-1, and SL acknowledges 
the NWO/STW, VICI grant 10828, and the DFG, project SPP1482 B12, for
partial financial support. Finally, we acknowledge the constructive criticism 
of the referees of the paper.

The final publication is available at Springer via http://dx.doi.org/10.1007/s10035-013-0430-z

\end{document}